\documentclass{aa}
\usepackage{graphics}

\begin{document}

\thesaurus{06(02.01.2; 02.09.1; 08.02.1; 08.14.2)}

\title{The zoo of dwarf novae : illumination, evaporation and disc radius
variation}
\titlerunning{The zoo of dwarf novae}

\author{Jean-Marie Hameury\inst{1}\inst{2}
        \and Jean-Pierre Lasota\inst{2}\inst{3}
		\thanks{also: DARC, Observatoire de Paris, France}
        \and Brian Warner\inst{4}
}
\offprints{J.-M. Hameury}

\institute{UMR 7550 du CNRS, Observatoire de Strasbourg,
           11 rue de l'Universit\'e, F-67000 Strasbourg, France,
           hameury@astro.u-strasbg.fr
\and Institute for Theoretical Physics, University of California,
        Santa Barbara, CA 93106-4030, USA
\and Institut d'Astrophysique de Paris, 98bis Boulevard Arago, F-75014 Paris,
        France, lasota@iap.fr
\and Department of Astronomy, University of Cape Town, Private Bag,
        Rondebosch 7700, Cape Town, South Africa,
        warner@physci.uct.ac.za
}

\date{Received / Accepted}

\maketitle

\begin{abstract} In the framework of the dwarf nova thermal-viscous disc
instability model, we investigate the combined effects on the predicted dwarf
nova lightcurves of irradiating the accretion disc and the secondary star and
of evaporating the inner parts of the disc. We assume the standard values of
viscosity. We confirm the suggestion by Warner (\cite{w98}) that the large
variety of observed outbursts' behaviour may result from the interplay of these
three effects. We are able to reproduce light curves reminiscent of those of
systems such as \object{RZ LMi} or \object{EG Cnc}. We can obtain long lasting
outbursts, very similar to superoutbursts, without assuming the presence of a
tidal instability.

\end{abstract}

\section{Introduction}

Dwarf novae (DN) are cataclysmic variable binary systems which, every few
weeks, exhibit 4 -- 6 mag outbursts, which last for a few days (see e.g. Warner
\cite{w95}). In several subclasses of DN both the outburst durations and
recurrence times can be very different from the values quoted above. It is
generally accepted that DN outbursts are due to a ``thermal-viscous"
instability. This instability occurs in the accretion disc in which the
viscosity is given by the $\alpha$-prescription (Shakura \& Sunyaev
\cite{ss73}), at temperatures close to 8000 K. Hydrogen is then partially
ionized and opacities are a steep function of temperature (see Cannizzo
\cite{c93} for a review and Hameury et al. \cite{hmdlh98} for the most recent
version of the model). Modeling dwarf nova lightcurves requires a varying
Shakura-Sunyaev parameter $\alpha$ (Smak \cite{s84}): it must be of the order
of 0.1 -- 0.2 (0.2 according to Smak 1999b) in outburst and of the order of
0.01 in quiescence, when the temperature is below the hydrogen ionization
temperature. One could therefore expect that the disc instability model (DIM)
might offer useful constraints on mechanisms which generate accretion disc
viscosity. This assumes, of course, a successful application of the model to
the observed DN outburst cycles.

However, despite its success in explaining the overall characteristics of DNs,
the DIM in its standard version faces several serious difficulties when one
tries to account for the detailed properties of dwarf nova outbursts. Some of
these difficulties are the result of an incomplete version of the DIM. For
example it was believed that a truncation of the inner parts of the disc is
necessary to explain the long delay between the rise of optical light and that
of UV and EUV in systems such as SS Cyg. As shown by Smak (\cite{s98}),
however, when correct outer boundary condition are assumed, the standard DIM
reproduces the observed delays. On the other hand, observed quiescent X-ray
fluxes far exceed the predictions of the model and seem to require an inner
`hole' in the disk. Such a hole can either be due to evaporation of the disc
close to the white dwarf (Meyer \& Meyer-Hofmeister \cite{mm94}), or to the
presence of a magnetic field strong enough to disrupt the disc. In addition,
systems such as WZ Sge, which have long recurrence time and large amplitude,
long outbursts require very low values of $\alpha$ ($\alpha < 10^{-4}$) if
interpreted in the framework of the standard DIM (Smak \cite{s93}, Osaki
\cite{o95}, Meyer-Hofmeister et al. \cite{mml98}). These values, much lower
than those of other DNs at similar orbital periods, are, however, left
unexplained. On the other hand, WZ Sge systems can be explained with standard
values of $\alpha$, provided that the disc is truncated as in other systems so
that it is either stable or marginally unstable (Lasota et al. \cite{lhh95},
Warner et al. \cite{wlt96}) and the mass transfer from the secondary is
significantly increased during the outburst under the influence of illumination
by radiation from the accreting matter (Hameury et al. \cite{hlh97}).

SU UMa systems are a subclass of dwarf novae which occasionally show long
outbursts during which a light\-curve modulation (superhump) is observed at a
period slightly longer than the orbital period; these superoutbursts are,
usually, separated by several normal outbursts. The superhump is due to a 3:1
resonance in the disc which causes the disc to become eccentric and to precess
(Whitehurst \cite{w88}). Osaki (\cite{o89}) proposed that the related tidal
instability is also responsible for the long duration and large amplitude of
superoutbursts (see Osaki \cite{o96} for a review of the thermal-tidal
instability model). In his model, the tidal torques which remove angular
momentum from the outer parts of the disc are increased by approximately an
order of magnitude when the disc reaches the 3:1 resonance radius (typically
$0.46 a$, where $a$ is the orbital separation) until the disc outer radius has
shrunk to typically $0.35 a$, i.e. by about 30\%. This model accounts for many
of the properties of SU UMa systems; it has, however, difficulties in explaining
systems with very short superoutburst cycles such as \object{RZ LMi}, for which
one must assume that the tidal instability stops when the disc has shrunk by
less than 10\% (Osaki \cite{o95b}), i.e. by much less than assumed in the
standard case.

Finally, it must be noted that a number of systems exhibit very bizarre
lightcurves: we have already mentioned the case of RZ LMi, which has
similarities, with systems such as \object{ER UMa}, \object{V1159 Ori} and
\object{DI UMa}. The case of the December 1996 superoutburst of \object{EG Cnc}
which was followed by 6 closely spaced normal outbursts in 1996 has not been
reproduced by simulations, except by Osaki (\cite{o98}) who assumed that the
viscosity parameter $\alpha_{\rm c}$ in the cold state was increased to 0.1,
almost the value in the hot state, for 70 days after the superoutburst, and
then returned to its quiescent value $\alpha_{\rm c} = 0.001$. It is finally
worth mentioning that the prototypical classical dwarf nova \object{U Gem}
exhibited an unusually long outburst in 1985, lasting 45 days, with a shape
similar to superoutbursts, but without superhumps (Mattei et al. \cite{m87}).
Since in this system the radius of the 3:1 resonance is larger than the size of
the primary's Roche lobe, one can conclude that while superhumps can be
attributed to the 3:1 resonance, the tidal instability is obviously not the
sole cause of very long outbursts.

The inability of numerical models to reproduce the large variety of observed
light curves may indicate that additional physical effects should be added to
the DIM. One such effect is the tidal instability. Another important class of
effects is the illumination of the disc and the secondary star. These effects
are usually not included in simulations, and it was suggested by Warner
(\cite{w95c}) that irradiation of the secondary star, that gives rise to high
and low states of mass transfer, and of the inner disc, that drastically
affects the disc instability, account for the light curves of VY Scl
stars. He also suggested (Warner \cite{w98}) that the wide spectrum of
superoutburst behaviours is generated by the interplay of reactions of the disc
and the secondary to irradiation. This suggestion is supported by observations
since there is evidence that the mass transfer rate from the secondary star
increases during outbursts (Smak \cite{s95}) in dwarf novae such as \object{Z
Cha} or \object{U Gem}, most probably as a result of illumination of the
secondary.

Recently Smak (\cite{s99a}) argued that properties of outburst cycles of
``standard" U Gem--type dwarf novae, which cannot be reproduced by the DIM (e.g.
the same maximum brightness of narrow and wide outbursts), are well explained if
{\sl all} outbursts are associated with some mass--transfer enhancement due to
the secondary's irradiation. (Long outbursts would be due to important mass
transfer enhancements, making the mass transfer rate larger than the critical
value for stability). If this is the case the `pure' DIM would find no
application in the real world.

It also appears that illumination of the disc itself by the hot white dwarf has
strong effects on the stability properties of the disc as soon as the white
dwarf temperature exceeds 15,000 K (King \cite{k97}, Hameury at al.
\cite{hld99}).

In this paper, we investigate the influence of the combined effects of
illumination of the disc, of the secondary star and of evaporation of the inner
parts of the disc, on the predictions of the DIM in which standard values of
viscosity are assumed. Our free parameters are the white dwarf mass, the
quiescent white dwarf temperature, the mass transfer rate in the absence of
illumination, and two parameters describing in a crude manner evaporation
effects and the influence of illumination of mass transfer from the secondary.
We show that a large variety of light curves are predicted by the models, many
of which have an observational counterpart. In section 2, we show that the long
1985 outburst of U Gem requires enhanced mass transfer during the outburst. In
section 3, we describe the model and our assumptions; in section 4, we give our
results and compare them with lightcurves of observed systems, and we discuss
briefly possible extensions of this work.

\section{The long 1985 outburst of U Gem}

U Gem is a prototypical dwarf nova that undergoes outbursts that last for 7-14
days, with an average recurrence time of 120 days (Szkody \& Mattei
\cite{sm84}). The orbital period is 4.25 hr, and the primary and secondary
masses are respectively 1.1 and 0.5 M$_\odot$ (Ritter \& Kolb \cite{rk98}); for
such parameters, the outer disc radius, defined as the radius of the last
non-intersecting orbits, is 4.15 10$^{10}$ cm.

The 1985 outburst that lasted for 45 days is therefore exceptional; long
outbursts are observed in other systems (SS Cyg for example), and are a natural
outcome of models in which the outer disc radius is kept constant (see e.g.
Hameury at al. \cite{hmdlh98}), which may happen when the outer disc radius
reaches the tidal truncation radius. However, one does not normally obtain
extremely long outbursts; the 1985 outburst of U Gem was in fact so long that,
more mass was accreted during this outburst than what was contained in the
pre-outburst disc. The maximum mass $M_{\rm d,max}$ of a quiescent disc during
quiescence is the integral of the maximum surface density $\Sigma_{\rm max}$ on
the cool branch; using the fits given by Hameury et al. (\cite{hmdlh98}), one
gets 
\begin{equation} 
M_{\rm d,max} = 2 \times 10^{20} \alpha_{\rm c}^{-0.83}
M_1^{-0.38} \left( {r_{\rm out} \over 10^{10} \rm cm} \right)^{3.14}
\end{equation}
where $\alpha_{\rm c}$ is the Shakura-Sunyaev viscosity parameter on the cool
branch, $M_1$ is the primary mass in solar units, and $r_{\rm out}$ is the
outer disc radius.

On the other hand, during a long outburst, the whole disc is entirely on the
hot branch, and the local mass transfer rate in the outer regions of a disc
must be large enough to prevent a cooling wave formation; using again the
analytical fits of Hameury al al. (\cite{hmdlh98}), this gives
\begin{equation}
\dot{M}_{\rm out} > 8 \times 10^{15} M_1^{-0.89} \left( {r_{\rm out} \over 
10^{10} \rm cm}\right)^{2.67}
\end{equation}
Here, $\dot{M}_{\rm out}$ is the local mass transfer rate in the outer parts of the disc,
which is close to the mass accretion rate onto the white dwarf if the whole
disc sits for a long time on the hot branch.

The maximum duration of such an outburst is thus $t_{\rm max} = M_{\rm
d,max}/\dot{M}$. For the parameters appropriate for U Gem, one gets
\begin{equation}
t_{\rm max} = 26 \left( {\alpha_{\rm c} \over 0.01}\right) ^{-0.83} M_1^{0.51}
\left( {r_{\rm out} \over 4.1 \; 10^{10} \rm cm} \right)^{0.47} \; \rm d
\end{equation}
As $\alpha_{\rm c}$ is larger than 0.01 for this prototypical dwarf nova (Livio
\& Spruit \cite{ls91} for example find that $\alpha_{\rm c}$ must be equal to
0.044 to account for the timing properties of U Gem), $t_{rm \max}$ can never
be as high as 45 days. This means that the total amount of mass accreted during
this long outburst is larger than the mass of the disc in quiescence; this is
possible only if the mass transfer rate from the secondary has increased to a
value close to the mass accretion rate onto the white dwarf, i.e. is close to
the critical rate for stable accretion. Such an increase of the mass transfer
rate is very likely caused by the illumination of the secondary. This
reinforces the conclusion of Smak (\cite{s99a}) that long outbursts result from
large mass--transfer enhancements.

\section{The model}

\subsection{Disc irradiation}

We use here the numerical code described in Hameury et al. (\cite{hmdlh98}).
This code solves the usual mass, angular momentum and energy conservation
equations on an adaptive grid, with a fully implicit scheme. This allows to
resolve narrow structures in the accretion disc (Menou et al. \cite{mhs99}),
and avoids the Courant condition which would severely limit the time step. To
describe disc irradiation we use a version of the code described in Dubus et
al. (\cite{dlhc99}) (see also Hameury et al. \cite{hld99}). A grid of vertical
structures is used to determine the cooling rate of the disc as a function of
the vertical gravity, the integrated disc surface density $\Sigma$, the central
temperature $T_{\rm c}$ and the illumination temperature $T_{\rm ill}$, defined
as $T_{\rm ill} = (F_{\rm ill}/\sigma)^{1/4}$ where $F_{\rm ill}$ is the
illuminating flux. In what follows, we use $\alpha_{\rm hot} = 0.2$ and
$\alpha_{\rm cold} = 0.04$, except where otherwise stated. We also neglect the
albedo $\beta$ of the disc; taking it into account introduces a multiplicative
factor $(1-\beta)^{-1/4}$ for the white dwarf temperatures.

It must also be stressed that the white dwarf surface temperature cannot be
too large; this is because the intrinsic (quiescent) white dwarf luminosity
must be significantly less than the accretion luminosity in outburst; for $M_1$
= 0.6 M$_\odot$, the quiescent white dwarf temperature has to be smaller than
33,000 K if the outburst amplitude is to be larger than 2 magnitudes.

\subsection{Illumination of the secondary}

We are interested here in the effect of illumination of the secondary on its
mass transfer rate on short time scales (days), and we do not consider any long
term effects that may lead to cycles accounting for the observed dispersion of
the average mass transfer rate for a given orbital period (McCormick \& Frank
\cite{mf98}). Even on short time scales, the response of the secondary to
illumination is complex (see e.g. Hameury et al. \cite{hlk88}); we prefer to
use here a simpler approach in which we assume a linear relation between the
mass transfer rate from the secondary $\dot{M}_{\rm tr}$ and the mass accretion
rate onto the white dwarf $\dot{M}_{\rm acc}$, i.e.

\begin{equation}
\dot{M}_{\rm tr} = \max (\dot{M}_0, \gamma \dot{M}_{\rm acc})
\label{eq:ill_sec}
\end{equation}
where $\dot{M}_0$ is the mass transfer rate in the absence of illumination.
This is similar to the formula used by Augusteijn et al.
(\cite{aks93}) in the context of soft X-ray transients. Although it is an
extremely crude approximation, it has, nevertheless, the advantage of having
only one free parameter $\gamma$. Its value must be in the range $[0-1]$ for
stability reasons.

Such an approach obviously requires the illumination to have a noticeable
effects on the secondary's surface layers. As mentioned earlier, strongly
irradiated companion stars are observed in several systems. To describe the
effects of irradiation of a Roche-lobe filling star we shall follow the approach
of Osaki (\cite{o85}) and Hameury et al. (\cite{hkl86}).

The mass transfer rate from the secondary can be written as (Lubow \& Shu
\cite{ls75}):
\begin{equation}
\dot{M}_{\rm tr} = Q \rho_{L1} c_s
\end{equation}
where $Q$ is the effective cross section of the mass transfer throat at the
Lagrangian point $L_1$, $Q = 1.9 \times 10^{17} T_4 P_{\rm hr}^2$ cm$^2$, where
$T_4$ is the surface temperature of the secondary and $P_{\rm hr}$ the orbital
period in hours; $c_s$ is the sound speed, and $ \rho_{L1}$ the density at
$L_1$, which, in the case of an isothermal atmosphere, can be expressed as:
\begin{equation}
\rho_{L1} = \rho_0 e^{(R-R_{L1})/H}
\end{equation}
where $\rho_0$ is the density calculated at a reference level, $R$ the
secondary radius (defined by this reference position), $R_{L1}$ the Roche lobe
radius and $H$ the scale height.

In quiescence, the mass transfer rate is low, of order of $10^{15} - 10^{16}$
g~s$^{-1}$. At short orbital periods and correspondingly low secondary's
temperatures, both $Q$ and $c_s$ are small. If one assumed that $\rho_0$
corresponds to the photospheric density, which for a low mass main sequence
star is of the order of $10^{-5}$ g cm$^{-3}$, one would find $\dot{M}_{\rm tr}
\sim 10^{16} e^{(R-R_{L1})/H}$ g~s$^{-1}$. This in turn would mean that
$(R-R_{L1})/H$ is small so that $\dot{M}_{\rm tr}$ varies as $T_4^{3/2}$, and
therefore is not very sensitive to illumination. In the irradiated case,
however, the reference density $\rho_0$ does not correspond to the photospheric
density but to the density at the base of the isothermal atmosphere, which
extends much deeper than the unilluminated photosphere when the atmosphere is
illuminated with an irradiation temperature exceeding 10$^4$~K. For those high
illumination fluxes, the outer layers of the star are affected on a thermal
time scale (seconds) at least down to a point where the unperturbed temperature
equals the irradiation temperature. This point is the base of the isothermal
layers in the illuminated case, and from models of very low mass stars (Dorman
et al. \cite{dnc89}), one gets $\rho_0 \sim 10^{-3}$ g~cm$^{-3}$, with
resulting very high mass transfer rates, exceeding 10$^{18}$ g~s$^{-1}$.

The latter estimate is an upper limit, as it does not take into account the
fact that a fraction of the secondary is shielded from irradiation by the
accretion disc; partial shielding does not suppress the enhancement of mass
transfer, since circulation at the surface of the star prevents the existence of
large temperature gradients, but reduces it in a complex way which we are not
attempting to describe here.

Observations show, however, that the increase can be quite significant; the mass
transfer rate rises by a factor 2 in \object{U Gem} and \object{Z Cha} (Smak
\cite{s95}), whereas Vogt (\cite{v83}) finds that in \object{VW Hyi}, the
bright spot luminosity increases by a factor $\sim$ 15 during maximum and
decline of outbursts close to a superoutburst, which he attributed to a
corresponding increase in mass transfer under the effect of illumination.
Although the evidence is not very strong, there are some indications that the
hot spot brightening, and hence the mass transfer increase, is delayed with
respect to the eruption by a day or two; this could be either the response time
of the secondary (Smak \cite{s95}), or the thermal inertia of the white dwarf
(only an equatorial belt is instantaneously heated by accretion, so irradiation
of the secondary is delayed).

\subsection{Inner disc radius}

There is evidence that in dwarf novae accretion discs are truncated, as
indicated by emission line profiles (Mennikent \& Arenas \cite{ma98}), or the
detection of a significant quiescent X-ray and UV flux (Lasota \cite{l96}).
If this is due to the presence of a magnetic field, the inner disc radius
$r_{\rm in}$ is a simple function of the mass accretion rate onto the white
dwarf:
\begin{equation}
r_{\rm in} = 9.8 \times 10^8 \dot{M}_{15}^{-2/7} M_1^{-1/7} \mu_{30}^{4/7} \;
\rm cm
\label{eq:rin}
\end{equation}
where $\dot{M}_{15}$ is the mass accretion rate in units of 10$^{15}$
g~s$^{-1}$, $M_1$ is the white dwarf mass and $\mu_{30}$ is the magnetic moment
in units of 10$^{30}$ Gcm$^3$. The value of $\mu_{30}$ should be such as to
allow $r_{\rm in} = R_1$ in outbursts, where $R_1$ is the primary radius, as
most DNs do not show then coherent pulsations. In quiescence, however, coherent
oscillations are observed (Patterson et al. \cite{prkm98}), and for example in
\object{WZ Sge} $\mu_{30} \approx$ 50 (Lasota et al. \cite{lkc99}).

An inner hole in the disc can be also due to evaporation. The physics of
evaporation is poorly understood but several models were proposed (Meyer \&
Meyer-Hofmeister \cite{mm94}, Liu et al. \cite{lmm97}, Kato \& Nakamura
\cite{kn98}, Shaviv et al. \cite{sww99}). The evaporation rates $\dot{\Sigma}$
in the disc are, however, quite uncertain. Evaporation is normally accounted for
by introducing the additional term $\dot{\Sigma}$ in the mass conservation
equation. However, because evaporation is expected to increase towards the
accreting body, and since the local mass transfer rate in the disc increases
sharply with radius during quiescence, the effects of evaporation are important
essentially very close to the disc inner edge, and can be treated assuming that
the disc inner radius is a function of the accretion rate onto the white dwarf,
just as in the case of the formation of a magnetosphere. To first order, the
effect of evaporation is to create a hole in quiescence which increases the
recurrence time; what matters is thus the inner disc radius in quiescence and
not the detailed way $r_{\rm in}$ varies with $\dot{M}_{\rm acc}$.

We shall therefore use equation (\ref{eq:rin}) in all cases, using $\mu_{30}$
as a free, unconstrained parameter that merely describes the size of the hole
generated in the disc by either the presence of as magnetic field, or by
evaporation.

\section{Results}

\begin{figure}
\resizebox{\hsize}{!}{\includegraphics{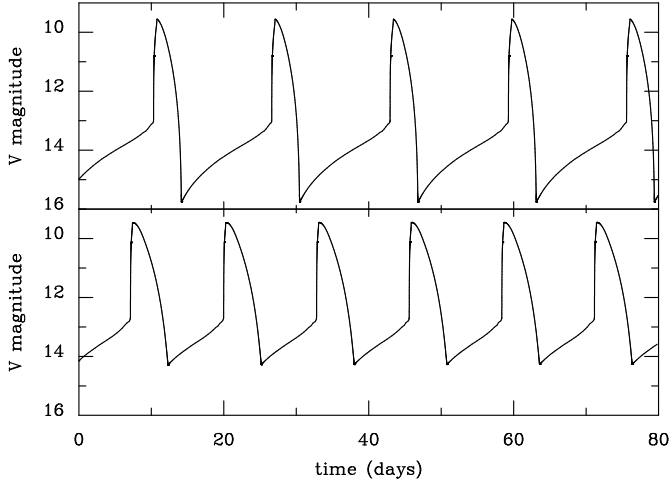}}
\caption{Predicted light curve in the standard model (no irradiation, fixed
inner radius) for an average mass transfer rate of $3 \times 10^{16}$
g~s$^{-1}$, an average disc radius of $1.9 \times 10^{10}$ cm, and a 0.6
M$_\odot$ (top panel), or 1.0 M$_\odot$ (bottom panel) primary.}
\label{fig:std}
\end{figure}

In the following, we discuss the influence of each individual effect mentioned
above; our reference situation is that of a system with a 1.0 M$_\odot$ primary,
whose radius is $5 \times 10^8$ cm; the mass transfer rate is $3 \times
10^{16}$ g~s$^{-1}$, and the average outer disc radius is $1.8 \times 10^{10}$
cm. These parameters are typical of short period dwarf novae with massive
primaries. The light curve corresponding to the standard version of the DIM is
given in Fig. (\ref{fig:std}). For comparison we also show the case of a system
with a 0.6 M$_\odot$ primary.

\subsection{Influence of the disc irradiation}

\begin{figure}
\resizebox{\hsize}{!}{\includegraphics{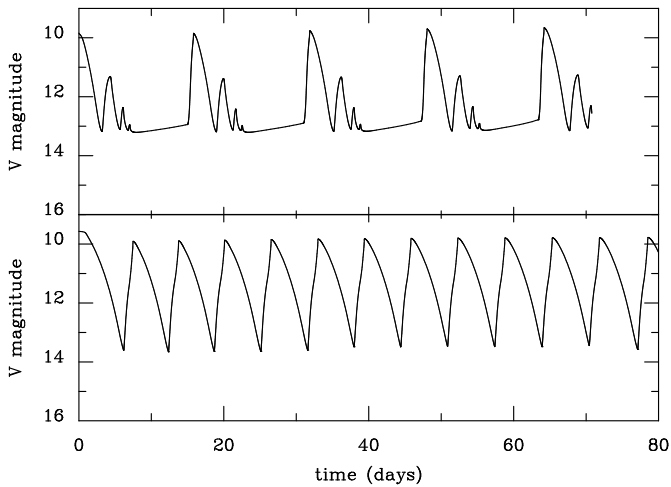}}
\caption{Predicted light curve when disc illumination is taken into account.
The parameters are the same as in Fig. \ref{fig:std}. The upper panel shows the
case of a 0.6 M$_\odot$ primary in which the radiation from the boundary layer
is ignored, the lower panel corresponds to a 1M$_\odot$ primary, and we have
taken into account the accretion luminosity. In both cases, the quiescent white
dwarf temperature is 30,000 K.}
\label{fig:ill_disc}
\end{figure}

The effects of irradiation of the disc by both the hot white dwarf and the
boundary layer has been described in detail in Hameury et al. (\cite{hld99}),
and we summarize here the most important results. For very hot white dwarfs
($T_{\rm eff} >$ 20~000 K), the temperature in the innermost parts of the disc
exceeds the hydrogen ionization temperature during quiescence; the viscosity is
therefore high in these regions, which are thus partially depleted as first
suggested by King (\cite{k97}). The transition region between the hot inner
disc and the outer, cool parts is strongly destabilized by irradiation, and the
model predicts several small outbursts between major ones. In particular, many
reflares are expected at the end of a large outburst (see Fig.
\ref{fig:ill_disc}). In certain cases, the reflares may dominated the light
curve; this depends on whether the heating front can reach the outer edge of
the disc or not. The reflares we obtained do not have the observed amplitudes
but it is tempting to attribute the succession of several normal outbursts
after a superoutburst in \object{EG Cnc} to this effect. Playing with
parameters would produce a result corresponding better to the observed
lightcurve but the merit of such an exercise is rather dubious considering the
important uncertainties of the model itself.

\subsection{Influence of the secondary irradiation}

Irradiation of the secondary enhances mass transfer. Hameury et al.
(\cite{hlh97}) showed that if one assumes that the effect of irradiation is
given by equation (\ref{eq:ill_sec}), outbursts having the general
characteristics of superoutbursts (long durations, flat top or exponential
decay with an abrupt cut-off) are expected. This model was, however, applied to a
case in which the quiescent mass transfer was low enough for the disc to be
stable on the cool branch; the instability was triggered by an external
perturbation of the mass transfer from the secondary. This was required to
explain the very long recurrence times of systems such as \object{WZ Sge} when
standard values of $\alpha$ are assumed. Marginally unstable mass transfer
rates as in Warner et al. (\cite{wlt96}) can give similar recurrence times but
also in this case the amount of mass accreted during the superoutburst requires
a substantial enhancement of mass transfer.

\begin{figure}
\resizebox{\hsize}{!}{\includegraphics{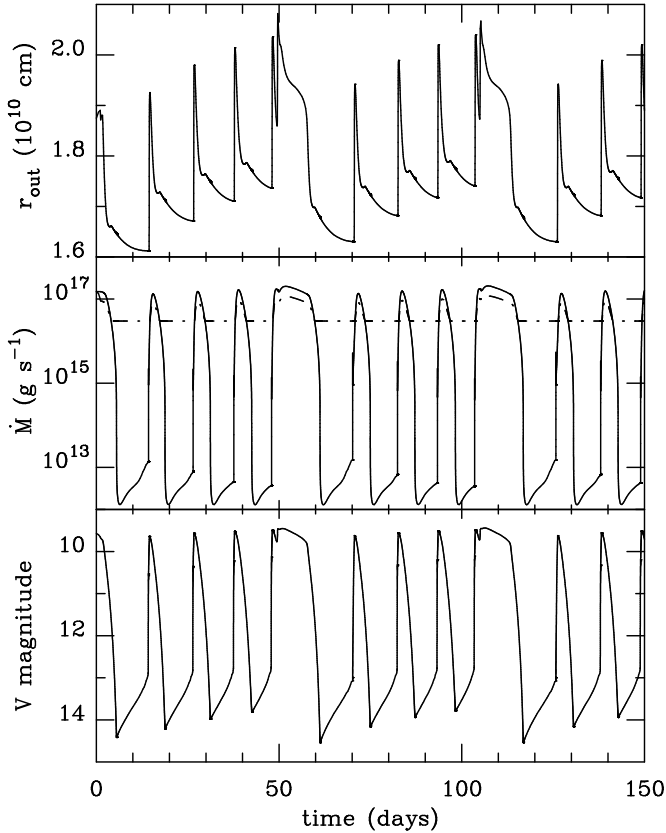}}
\caption{Effect of secondary illumination on the predicted light curve; the
parameters are those of Fig. \ref{fig:std}, with $M_1$ = 0.6 M$_\odot$ and
$\gamma$ = 0.5. The upper panel shows the outer disc radius, the middle panel
shows the accretion rate onto the white dwarf (solid line) and the mass
transfer rate from the secondary (dashed line); the lower panel shows the
visual magnitude of the disc.}
\label{fig:unill_m10}
\end{figure}

Figure \ref{fig:unill_m10} shows the light curve obtained when one includes the
secondary irradiation in the model. We neglect here disc irradiation and the
disc is not truncated. We have taken $\gamma$ = 0.5, and all other parameters
are as in Fig. (\ref{fig:std}), for a 1 M$_\odot$ primary (i.e. $\dot{M}_0 = 3
\times 10^{16}$ g s$^{-1}$). The light curve is similar to those observed in SU
UMa systems; it shows several normal outbursts separated by a large one which
is sustained by enhanced mass transfer from the secondary.

Large outbursts occur when the surface density at the outer edge of the disc is
large enough that a cooling wave does not start immediately after the heating
wave has arrived; equivalently, the disc mass must be larger than some critical
value, and one therefore expects that the recurrence time of such large
outbursts $T_{\rm s}$ varies roughly as $\dot{M_0}^{-1}$. This, however,
requires that the disc mass keeps increasing despite the presence of small
outbursts, which means that $\dot{M}_0$ must be large enough to refill the disc
with more mass than is lost during such outbursts. For $\dot{M}_0 = 3 \times
10^{15}$ g s$^{-1}$, which is more appropriate for short period systems, one
does not get long outbursts, at least for the value of $\alpha$ considered
here. Short outbursts are (as all our outbursts) of the inside-out type, and
their recurrence time $T_{\rm n}$ is the viscous time, and therefore do not
depend on the mass transfer rate. 

The correlation between $T_{\rm s} \propto T_{\rm n}^{0.5}$ found in SU UMa
systems (Warner \cite{w95b}) is interpreted, in the framework of the
tidal-thermal instability (TTI) model, as resulting from $T_{\rm s}$ varying as
$\dot{M}^{-1}$, as in our case, and from $T_{\rm  n} \propto \dot{M}^{-2}$ for
outside-in outbursts (Osaki \cite{o95}). One must however be careful with such
a simple interpretation. This explanation is valid only if $\dot{M}$ is the
only parameter determining both $T_{\rm n}$ and $T_{\rm s}$; this is clearly not
the case, as quantities such as the viscosity in quiescence (that may vary by
orders of magnitude from WZ Sge type systems to ``normal" SU UMa's), the disc
radius (to the power 5.6), and the orbital period enter together with $\dot{M}$
in expressions for $T_{\rm n}$ and $T_{\rm s}$ (Osaki \cite{o95}). It must also
be stressed out that, for the low mass transfer rates of SU UMa stars,
outside-in outbursts are not a natural outcome of the models and are produced
by lowering the value of $\alpha_{\rm c}$ or by making it an appropriate
function of radius.

The time evolution of the outer disc radius is different from the predictions
of the TTI model in several respects: (i) $r_{\rm out}$ varies during a normal
outburst, whereas in the TTI model, $r_{\rm out}$ remains roughly constant
during an outburst after an initial increase during the rise (ii) $r_{\rm out}$
oscillates at the beginning of a large outburst, (iii) the disk extends to a
larger radius during a large outburst than during a short outburst, allowing
for the possibility of the development of superhumps if the radius can reach
the 3:1 resonance radius, whereas in the TTI model the disk size varies by 30\%
during a superoutburst, with an average that is smaller than the average size
during the previous normal outburst, and (iv) $r_{\rm out}$ remains
approximately constant during a superoutburst, showing only a slow decline.

Our results are similar to those of Smak (\cite{s91b}), who considered the
effect of enhanced mass transfer during superoutbursts, and concluded that
observed disc radius variations in \object{Z Cha} and the length of the cycle
in \object{VW Hydri} appeared to support the enhanced mass transfer model. The
main difference with our work comes from the approximations describing the
effect of illumination: whereas we assume a dependence between $\dot{M}_{\rm
tr}$ and $\dot{M}_{\rm acc}$ given by Eq. \ref{eq:ill_sec}, in Smak's model
$\dot{M}_{\rm tr}$ is increased by about one order of magnitude after the
maximum of a normal outburst, during a fixed period. As a consequence, the disc
radius we obtain at the end of a large outburst is much smaller than in Smak's
model. In our case, when the cooling wave starts propagating, the accretion
rate onto the white dwarf, and hence the mass transfer from the secondary, is
unaffected, and the disc contracts as in the unilluminated case, whereas in
Smak's model, the disk expands rapidly when mass transfer is reduced by a
factor 10; the surface density then drops at the outer edge below the critical
value, and a cooling wave starts in quite a large disc.

Ichikawa et al. (\cite{iho93}) also considered a mass transfer outburst as the
source of superoutbursts. They compared the resulting lightcurves with those
produced by the tidal--thermal model and concluded in this last model gives a
much better representation of observed properties of SU UMa's system. One
should stress, however, that in Ichikawa et al. (\cite{iho93}) superoutbursts
are triggered by a mass transfer `instability'. In our case, superoutbursts are
triggered by the usual thermal-viscous instability and it is only the
subsequent evolution of the outburst which is modified by an enhanced mass
transfer. In Ichikawa et al. (\cite{iho93}), the mass transfer $\dot{M}_{\rm
tr}$ is increased by a factor 100, whereas in our case the peak mass transfer
rate from the secondary is $1.2 \times 10^{17}$ g~s$^{-1}$, i.e. increased only
by a factor 4. It is much smaller than the maximum possible rate from an
irradiated low-mass star (see Section 2.1). Note that the average mass transfer
rate, as given by Eq. (\ref{eq:ill_sec}) is larger than $\dot{M}_0$, and thus
larger than for Fig. \ref{fig:std}; it is in this case $4.8 \times 10^{16}$
g~s$^{-1}$.

Observations are, for the moment, not of much help in deciding which of the
models is right. There are good reasons to believe that a tidal instability is
required to account for the superhump phenomena. What is not known, however, is
(i) the increase in the tidal torque resulting from this instability, and (ii)
the radius at which the instability stops. There is also good evidence for an
enhanced mass transfer, due to irradiation, during outbursts but a reliable
description of this effect is missing. In both models, however, the observed
correlation with superoutburst and normal outburst frequency can be obtained
only by playing with the viscosity prescription which, of course, is not very
satisfactory.

The quiescent luminosities and accretion rates are almost identical in
the standard and enhanced mass transfer cases, as expected, since after
the passage of the cooling wave, the disc has essentially forgotten its
initial conditions; differences arise only from the different disk sizes
which are smaller in the illuminated case because of the large mass
transfer increase.

\subsection{Parameter dependence}

\subsubsection{Mass transfer rate}

In this section we shall consider the combined effect of irradiation of both
the disc and the secondary and shall determine how the resulting lightcurves
depend on the value of rate $\dot{M}_0$ at which mass is transferred from an
unilluminated secondary.

\begin{figure}
\resizebox{\hsize}{!}{\includegraphics{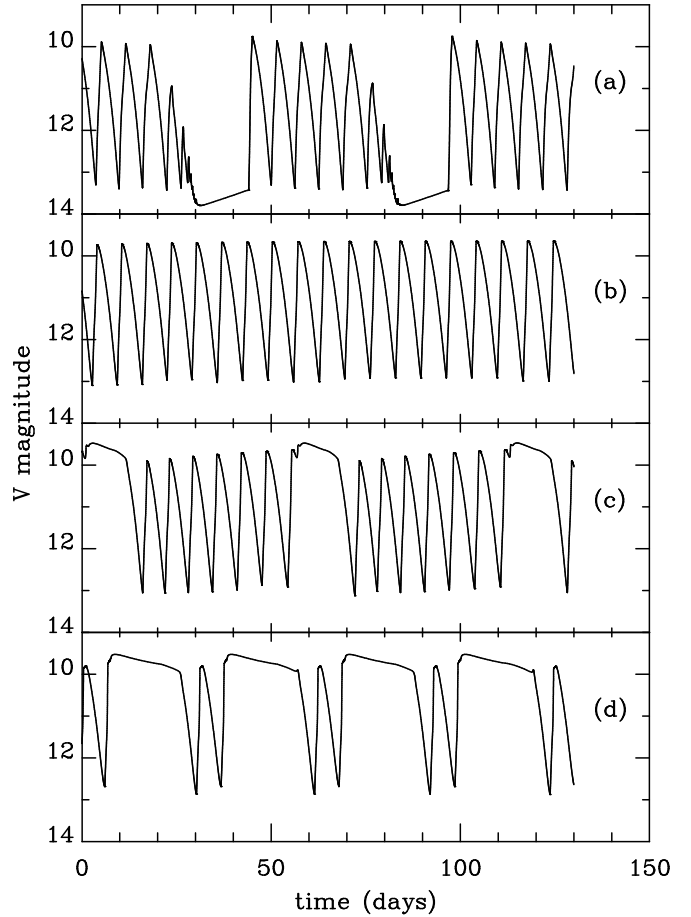}}
\caption{Visual magnitudes obtained in the case of a system containing a 1
M$_\odot$ primary, whose radius is $5 \times 10^8$ cm and surface temperature
35,000 K in quiescence. $\gamma$ = 0.5 for all four panels, and $\dot{M}_0$ =
$10^{16}$, $3 \times 10^{16}$, $4 \times 10^{16}$, and $7 \times 10^{16}$
g~s$^{-1}$ from top to bottom. Note that the average mass transfer from the
secondary is larger than $\dot{M}_0$.}
\label{fig:mdot}
\end{figure}

Figure \ref{fig:mdot} shows various light curves obtained by varying
$\dot{M}_0$. We considered a system containing a 1 M$_\odot$ primary, whose
radius is $5 \times 10^8$ cm and surface temperature 35,000 K in quiescence. We
have taken $\gamma$ = 0.5, and $\dot{M}_0$ ranges from $10^{16}$ to $7 \times
10^{16}$ g~s$^{-1}$; this corresponds to average transfer rates in the range
1.5 -- 8.6 $\times 10^{16}$ g~s$^{-1}$. The inner disc radius is equal to the
white dwarf radius. The white dwarf temperature has been chosen in the upper
range of observed values in order to emphasize illumination effects. The effect
is quite dramatic. Light--curves corresponding to cases (a) and (b) do not seem
to be observed (as mentioned in the previous section the mass transfer rate is
too low for long outburts to be present). It might be that the corresponding
systems exist, but have not yet been discovered because they are intrinsically
rare -- the parameters of Fig. \ref{fig:mdot} are at the upper range of allowed
values, or that some of the curves we obtain are artifacts due to our
oversimplified treatment of the secondary response to illumination. Lightcurves
(c) and (d), however, compare very well with those of systems having very short
supercycles such as \object{RZ LMi}. In our model they are obtained for high
mass transfer rates, which is natural since these systems spend most of their
time in the high state, whereas in the tidal--thermal instability such
lightcurves require an ad hoc reduction of the parameter describing the tidal
interaction. The agreement with \object{RZ LMi} can be improved; in particular
a reduction of the duration of the long outburst will be obtained by decreasing
$\gamma$.

One should note, however, that we have assumed that $\dot{M}_{\rm tr}$ responds
immediately to changes in $\dot{M}_{\rm acc}$, whereas one could argue that
there is a delay of the order of one or two days between illumination and the
increase of mass transfer as discussed above. The introduction of such a delay
makes the occurrence of long outbursts more difficult, as these require a near
balance between mass transfer and accretion onto the white dwarf that must be
established within a short outburst. We checked that if we use in Eq.
(\ref{eq:ill_sec}) the average of the mass accretion rate over the past 2.5
days (the duration of short outbursts), long outbursts are suppressed.

Finally, it is worth pointing out that small outbursts in Fig. \ref{fig:mdot}
are intrinsically different from those obtained when the disc illumination is
not taken into account. Whereas in a non--irradiated disc small amplitude
outbursts appear when the heating front cannot bring the whole disc into a hot
state, here the disc never returns to quiescence in its inner parts, so the
cooling wave is reflected into a heating wave when it gets close to the stable
hot inner part of the disc. The amplitude of these reflares grows as a
consequence of the enhanced mass transfer during maximum, until the disc mass
has grown up to a point where a self-sustained long outburst is possible.

\subsubsection{Viscosity}

\begin{figure}
\resizebox{\hsize}{!}{\includegraphics{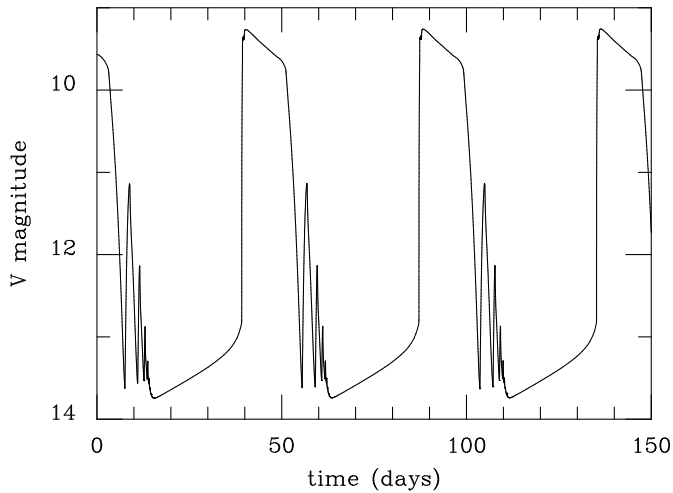}}
\caption{Light curves obtained with $\alpha_{\rm cold} = 0.02$; all other
parameters are the same as for Fig. \ref{fig:mdot}b.}
\label{fig:alpha}
\end{figure}

The reflares properties also depend on the ratio $\alpha_{\rm hot}/\alpha_{\rm
cold}$~: the smaller this ratio, the more important the reflares (see Menou et
al. \cite{mhln99} for a discussion of this effect in the context of X-ray
transients). This is simply due to the fact that, the lower this ratio, the
larger $\Sigma / \Sigma_{\rm max}$ after the passage of a cooling front,
$\Sigma_{\rm max}$ being the maximum surface density on the cold stable branch;
in the limiting case $\alpha_{\rm hot} = \alpha_{\rm cold}$ there are no
outbursts (Smak \cite{s84}), but a heating/cooling wave that propagates back
and forth. Fig. \ref{fig:alpha} shows the effect of changing $\alpha_{\rm
cold}$ to a smaller (by a factor 2) value. Successive reflares no longer reach
the outer edge of the disc, and their amplitude therefore decreases from one
mini-outburst to the next one. This accounts for the presence of flat top
outbursts which were absent in Fig. \ref{fig:mdot}b. The lightcurves are
similar for all mass transfer rates, showing the pattern of Fig \ref{fig:alpha}
with longer recurrence times for smaller $\dot{M}_{\rm tr}$. The only exception
is for $7 \times 10^{16}$ g~s$^{-1}$, which is close to stability, and for
which the main outbursts are of the outside-in type.

\subsubsection{Outer disc radius}

\begin{figure}
\resizebox{\hsize}{!}{\includegraphics{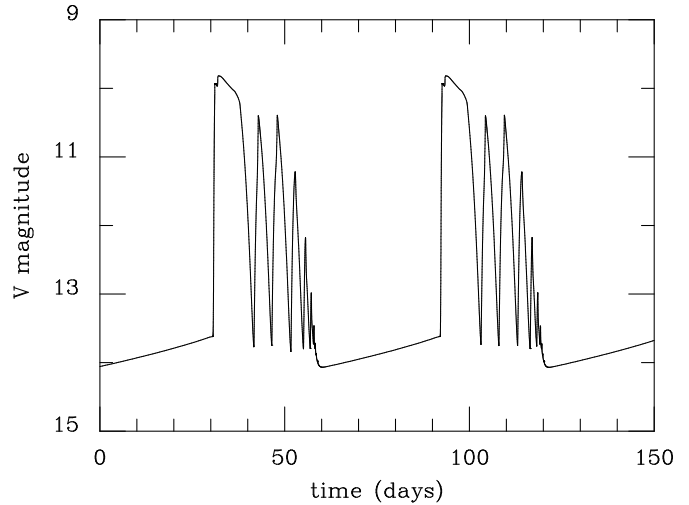}}
\caption{Example of a lightcurve obtained for a small disk; we consider a 1
M$_\odot$ primary with surface temperature 35,000 K, $\alpha_{\rm cold} = 0.02$,
$\alpha_{\rm hot}$ = 0.2, $\gamma = 0.5$, and $\dot{M}_0 = 10^{16}$ g~s$^{-1}$.}
\label{fig:rout}
\end{figure}

Since small discs favour large reflares, it is not surprising that when one
considers discs with average $r_{\rm out} = 1.3 \times 10^{10}$ cm, and one
takes $\alpha_{\rm cold}$ = 0.02 and $\alpha_{\rm hot}$ = 0.2, one obtains a
combination of the lightcurves shown in the two previous sections; Fig
\ref{fig:rout} is a good example of this. It is worth noting that such a light
curve is reminiscent of that of \object{EG Cnc}, even though the timescales are
not quite the same. We do obtain the right pattern for the reflares, but we do
not reproduce the very long superoutburst of \object{EG Cnc} (100 days), that
would require $\gamma$ to be very close to unity, meaning that the linear
approximation in Eq. \ref{eq:ill_sec} is invalid. For \object{WZ Sge}, one
already had to assume a relatively large value of $\gamma$ (0.87) in order to
reproduce the observed 25 days duration; since the outburst duration varies as
$1/\log(\gamma)$ (Hameury et al. \cite{hlh97}), we would need $\gamma = 0.97$
to obtain 100 days. Another difference with EG Cnc is the amplitude of the
minioutbursts: the observed ones have approximately the same amplitude, whereas
we get two identical minioutbursts, the others being of decreasing amplitude.
We have not been able to reproduce this behaviour with our parameterization; a
possible solution is to introduce a time dependent temperature of the white
dwarf. This is expected, because the superoutburst lasted long enough to heat
up the surface of the white dwarf that will then cool.

If the rebrightenings of \object{EG Cnc} are indeed due to illumination
effects, this implies that $\alpha_{\rm cold}$ cannot be small as we do not
obtain reflares when $\alpha_{\rm cold}$ is significantly less than 0.01; Osaki
et al. (\cite{ost97}) reached the same conclusion, but on different grounds;
they assumed that $\alpha_{\rm cold}$ was increased to 0.1 during the
superoutburst, remained high for 2 months, and then decreased back to small
values (0.001), and had therefore to set $\alpha_{\rm cold}$ to be an explicit
function of time.

\subsubsection{Inner disc radius}

Finally we consider the effect of removing the inner disc regions, keeping both
the disc and secondary irradiated.

\begin{figure}
\resizebox{\hsize}{!}{\includegraphics{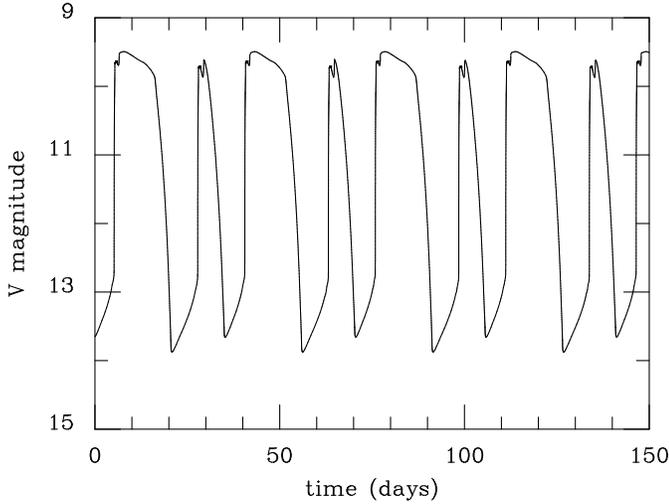}}
\caption{Lightcurve obtained when the inner disc is truncated at a radius given
by Eq. (\ref{eq:rin}), with a magnetic moment of $2 \times 10^{30}$ G~cm$^3$,
the other parameters being that of Fig. \ref{fig:mdot}c.}
\label{fig:rin}
\end{figure}

Apart from increasing the delay between the onset of an outburst in the disc
and accretion onto the white dwarf, a large inner disc radius has a
stabilizing effect on the disc itself, by preventing inside-out outbursts.
This effect is quite noticeable in the case where the white dwarf surface
temperature is high: if the inner disc radius $r_{\rm in}$ is large enough,
the unstable transition region between the stable region heated above
hydrogen ionization temperature and the cooler, quiescent external part
does not exist. This suppresses the bounces after a longer outburst, as can
be seen in figure \ref{fig:rin}.

For a given mass transfer rate, there is a critical value of the inner radius
above which the disc is stable; when $r_{\rm in}$ approaches this value, the
recurrence rate goes to infinity. Arbitrarily long recurrence rates could
therefore be expected, but only at the expense of very fine tuning; in the
normal case where the mass accretion rate onto the white dwarf is negligible
in quiescence as compared with the mass transfer rate from the secondary, the
reasoning used by (Smak \cite{s93}) applies. The recurrence time $t_{\rm
rec}$ is equal to:
\begin{equation}
t_{\rm rec} = {\Delta M \over \dot{M}_{\rm tr}} = f {M_{\rm crit} \over \dot{M}_{\rm tr}}
\end{equation}
where $f$ is the ratio of the amount of mass transferred during an outburst
$\Delta M$ and the maximum possible disc mass $M_{\rm crit}$, obtained
assuming that the surface density is everywhere the critical surface density.
Numerical models show that the surface density is not very far from its
critical value even at large radii, and that the amount of mass transferred
during a normal outburst is typically 10\% of the total disc mass.
Therefore, $f$ is not a very small parameter that could freely vary, and
large changes in $t_{\rm rec}$ cannot result from variations in $r_{\rm in}$
alone. A similar situation is encountered in the case of soft X-ray
transients (Menou et al. \cite{mhln99}).

\section{Conclusions}

We have shown that many types of light curves can be produced by numerical
models that include the illumination of both the secondary and the accretion
disc, thereby explaining a great variety of observed light curves. These
effects account for phenomena such as post-outburst rebrightening (e.g.
\object{EG Cnc}), long outbursts (\object{U Gem} for example), or SU UMa
systems with extremely short supercycles. In order to explore further these
possibilities, one would need to determine from observations the mass transfer
rate from the secondary as a function of the mass accretion rate onto the white
dwarf, with a better accuracy than it is available now.

Despite the fact that our approximations are very crude, in particular the one
concerning the response of the secondary to illumination, we can nevertheless
draw a number of conclusions. First, the illumination of the disc is important
only if the white dwarf is relatively massive, so that it can have a high
temperature without contributing too much to the light emitted by the system in
outburst, and that the efficiency of accretion is high. Rebrightenings also
require $\alpha$ not to be too low in quiescence.

The fact that we can reproduce an alternance of normal and long outbursts when
the illumination of the secondary is included does not of course imply that the
thermal-tidal instability model for SU UMa is incorrect; a tidal instability is
most probably required to account for the superhump phenomenon. The
question of the precise role of this instability is, however, still open and our
results raise some doubt on the validity of the parameters derived when fitting
the observations, in particular for systems having a very short supercycle.

This also means that the determination of the viscosity from the modeling
of light curves is a far more difficult task than previously estimated. We
obviously need some progress in the determination of the tidal torque; we also
need to know how the secondary responds to illumination. We finally should
include 2D effects in our models. First because the orbits in the outer disc
are far from being circular, and second because the presence of a hot spot
whose temperature can be of order of 10,000 K could in principle significantly
alter the stability properties of the outer disc.

\begin{acknowledgements}
We thank the referee, Professor Y. Osaki, for helpful comments and criticisms.
This research was supported in part by the National Science Foundation Grant
No. PHY94-07194.
\end{acknowledgements}

\end{document}